\documentclass{article}
\usepackage{smc2017}
\usepackage{times}
\usepackage{ifpdf}
\usepackage[english]{babel}
\usepackage{cite}
\usepackage{algorithm}  
\usepackage{algorithmic}
\def\papertitle{Automatic Music Accompanist}
\def\firstauthor{First author}
\def\secondauthor{Second author}
\def\thirdauthor{Third author}

\newif\ifpdf
\ifx\pdfoutput\relax
\else
   \ifcase\pdfoutput
      \pdffalse
   \else
      \pdftrue
\fi

\ifpdf % compiling with pdflatex
  \usepackage[pdftex,
    pdftitle={\papertitle},
    pdfauthor={\firstauthor, \secondauthor, \thirdauthor},
    bookmarksnumbered, % use section numbers with bookmarks
    pdfstartview=XYZ % start with zoom=100% instead of full screen; 
   ]{hyperref}
  \usepackage[pdftex]{graphicx}
  \graphicspath{{./figures/}}
  \DeclareGraphicsExtensions{.pdf,.jpeg,.png}

  \usepackage[figure,table]{hypcap}

\else % compiling with latex
  \usepackage[dvips,
    bookmarksnumbered, % use section numbers with bookmarks
    pdfstartview=XYZ % start with zoom=100% instead of full screen
  ]{hyperref}  % hyperrefs are active in the pdf file after conversion

  \usepackage[dvips]{epsfig,graphicx}
  \graphicspath{{./figures/}}
  \DeclareGraphicsExtensions{.eps}

  \usepackage[figure,table]{hypcap}
\fi

\hypersetup{
    colorlinks,%
    citecolor=black,%
    filecolor=black,%
    linkcolor=black,%
    urlcolor=black
}

\title{\papertitle}

 \twoauthors
   {Anyi Rao} {Nanjing University \\ %
     {\tt \href{mailto:anyirao@smail.nju.edu.cn}{anyirao@smail.nju.edu.cn}}}
   {Francis Lau} {The University of Hong Kong \\ %
     {\tt \href{mailto:fcmlau@cs.hku.hk}{fcmlau@cs.hku.hk}}}

\begin{document}
\capstartfalse
\maketitle
\capstarttrue
\begin{abstract}
Automatic musical accompaniment is where a human musician is accompanied by a computer musician. The computer musician is able to produce musical accompaniment that relates musically to the human performance. The accompaniment should follow the performance using observations of the notes they are playing.

This paper describes a complete and detailed construction of a score following and accompanying system using Hidden Markov Models (HMMs). It details how to train a score HMM, how to deal with polyphonic input, how this HMM work when following score, how to build up a musical accompanist. It proposes a new parallel hidden Markov model for score following and a fast decoding algorithm to deal with performance errors.
\end{abstract}

\section{Introduction}\label{sec:introduction}
Accompanists may not always be available when needed,
or available accompanists may not have sufficient
technical ability to provide adequate accompaniment.
A solution for many musicians is to make use of recorded
or computer-generated accompaniment where the
accompaniment is static, i.e. never changing from one
performance to another. This forces the musician to
adapt their playing to synchronize with the accompaniment.
It is more natural for the musician, though, if the
accompaniment adapts to the performer, particularly as
a musician’s playing tends to be 'free'.

To dynamically synchronize the accompaniment with
the performance by the musician, the accompanist
should track the performer’s progress through the score
of the piece as they play. Score following is the process
whereby a musician follows another musician’s playing
of a musical piece, by tracking their progress through the
score of that piece. The term is most commonly used in
the context of computer-generated accompaniment,
where one or more of the musicians involved are artificial
rather than human. The purpose of the research outlined
in this paper is to construct a automatic accompaniment.

In live performance, score following must be on-line real-time, i.e. producing accompaniment in time with the soloist’s playing. This
places extra challenges for the score follower. The system has a
more limited amount of information available for analysis: only the notes that have been played so far, as opposed to having the whole performance to analyse. It requires fast computation speed. The accompanist needs finish one accompaniment before the next note comes.

Our contributions are as follows,
\begin{itemize}
\item[1]Our work is the first free open-source Windows based automatic music follower and accompanist to our best knowledge.
\item[2]We construct a comprehensive system and show how it works with detailed theoretical induction, including score follower training/decoding and score accompanist. 
\item[3]We propose a fast decoding algorithm, reduced computational complexity from $O(n^2)$ down to $O(n)$. It is able to work in real time with practical length scores. 
\item[4]We build up two hands parallel HMM to improve accuracy and computational speed. 
\end{itemize}

\subsection*{Background}
There are several reasons why a musician may not perform the piece exactly as written.
Changes may be added by mistake:
1. A wrong note is played; 2. Extra notes are added; 3. Scored notes are missed out; 4. The musician loses their place in the music or starts playing from the wrong point in the score. 5. The tempo speeds up or slows down unintentionally.
Also changes may be added deliberately, as the musician adds their own interpretations to the music:
1. The musician adds embellishments such as trills, to ‘decorate’ the notes;
2. The tempo speeds up or slows down deliberately, for musical effect;
3. The piece being played may have rubato or free/improvised sections, where the musician is free to vary the tempo and notes played according to their own choice.

\subsection*{Terms}

• Performance: In the context of this project, a performance is defined specifically
as the situation where a solo musician (soloist), such as a flute player or
singer, performs a piece of music. The solo musician would be accompanied by
another musician (accompanist) on an instrument such as piano. This may be
in a concert or similar scenario, performing to an audience, but this condition is
not mandatory. What is important is that the soloist is making an attempt to play
through the piece in a linear fashion, from start to finish.

• Performer/Soloist: The solo musician who is performing the piece; what they
play is the most important part of the performance for any audience that may be
listening.

• Accompanist: The musician who is playing the accompaniment; supporting
the soloist’s performance.

• Melody/Solo melody: The music that is being played by the soloist.

• Accompaniment: The music which is played by an accompanist, during the
performance of the soloist. Accompaniment can be thought of background
music which is designed to enhance what the soloist is playing and support the
soloist’s performance.

• Score follower: A computer accompanist that follows the solo melody through
the score as it is being played, to produce accompaniment relative to where the soloist is in the score.

\section{Hidden Markov Models}

A musical score is divided up into a sequence of musical events. (for example one note or one beat can be considered as one modellable musical event)

The score follower uses a Hidden Markov Model to represent these musical events, and uses a decoding algorithm to estimate what state the performer is most likely to be in at that time, i.e. which musical event in the score the performer is currently playing.

\subsection{HMM Structure}
\begin{figure}[ht]
\centering
\includegraphics[width=0.9\columnwidth]{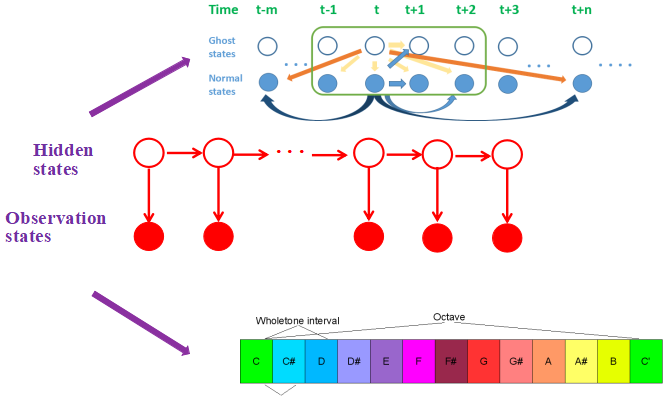}
\caption{Hidden Markov
Model structure.\label{HMM}}
\end{figure}

We define the observation states as 12 notes in the western musical chromatic scale. We ignores octave differences between notes and merely consider 12 possible observations:
$\lbrace$ C, C$\sharp$, D, E$b$, E, F, F$\sharp$, G, G$\sharp$, A, Bb, B $ \rbrace$, as shown in Figure \ref{HMM}. The hidden states base on beat and encode the information relates to the beat as detailed in the below section \ref{beatbased}. The paraments $\lambda$ of the HMM contains three parts $\lbrace$ $\pi , A , B$ $\rbrace$ denoted below.

\begin{itemize}
  \item N: the number of possible states. We use N symbols $S_1, S_2, \cdots , S_N$ to denote them.
  \item M:  the number of possible observations. 
  \item  \textbf{ $\pi$} : the prior (initial) state distribution. \\ $ \pi = (\pi_1, \pi_2, \cdots , \pi_N )$ and $\pi_i$ = Pr($Q_1 = S_i$).
  \item A: the state transition matrix. $A_{ij} = Pr(Q_t = S_j |Q_{t-1} = S_i) $
  is the probability of the next state being $S_j$ if the current one is $S_i$, 1 $\leq$ i, j $\leq$ N.
Note that A does not change when t changes.
  \item B: the observation probability matrix. Instead of denoting one probability
as $B_jk$, we use $b_j (k) = Pr(O_t = V_k|Q_t = S_j )$ to denote the probability of
the observation being $V_k$ when the state is $S_j$ , 1 $\leq$ j $\leq$ N, 1 $\leq$ k $\leq$ M.
And, B does not change when t changes.
\end{itemize}

\subsubsection{beat-based representation}\label{beatbased}
If there is a simple tune for which each note is of the same length, the naive choice is to model each note as an individual HMM state.

But when music pieces become more complex, it is no longer realistic to model each note as a new state, and instead the more pertinent aspect to model as a state is each beat, or a fraction of each beat. For such cases, it was necessary to consider how the timing information within the score should be modelled (in addition to how the notes should be modelled).

The two obvious ways to model a note that is held for longer than one state (i.e. notes that extend over a beat or more) are:
\begin{itemize}
\item[1.] Allow states with self-transitions, so the HMM stays in a given state while a note is being held and only moves out of that state when the note is released.
\item[2.] Have a finite number of states representing each note that is longer than one state, proportional to the length of the note (for example if each state represents one beat and a note is three beats long, represent it as three sequential states).
\end{itemize}

The more successful option here is the second \cite{anna2009} with more flexibility to vary the accompaniment and it also able to encode notes of different lengths into the HMM.

\subsubsection{Errors representation} 

There are three classes of probable errors \cite{orio2001score}:

\begin{itemize}
\item WRONG: An incorrect note is played in place of the correct note.

\item SKIP: A note in the score is missed out altogether.

\item EXTRA: An extra, unscored note is added in the performance.
\end{itemize}

\begin{figure}[ht]
\centering
\includegraphics[width=0.9\columnwidth]{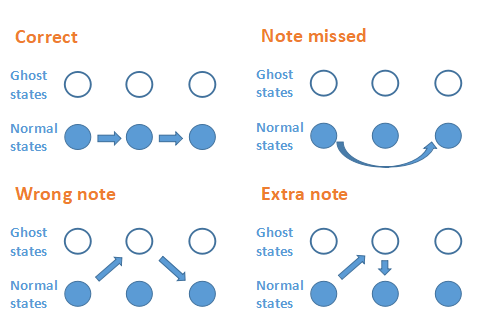}
\caption{Typical deviations from a score and the HMM hidden state transitions associated with these deviations.\label{errors}}
\end{figure}

The Hidden Markov Model processes such errors by the soloist, as they happen, by taking a specific path through the normal and ghost states. The paths for each class of error are shown in Figure \ref{errors}.

\begin{figure}[ht]
\centering
\includegraphics[width=0.9\columnwidth]{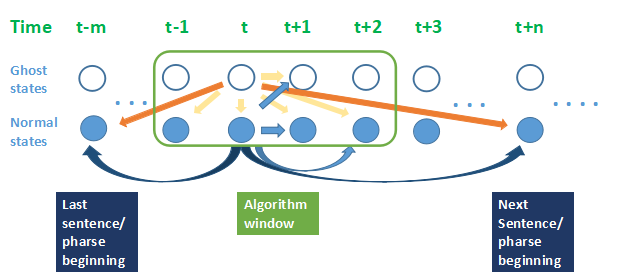}
\caption{All allowed transitions from the first normal/ghost state
pair.\label{alltransitions}}
\end{figure}

\subsection{Training}
We train the HMM involving getting the maximum probability of being in the correct normal state or ghost state, given a sequence of observations. We define four variables. 

\begin{itemize}
  \item $\alpha_t(i)$: $\alpha_t(i)= Pr(o_{1:t}, Q_t = S_i|\lambda) $ with recursion:
 $ \alpha_{t+1}(i) = (\sum_{j=1}^N=\alpha_t(j)A_{ji})b_j (o_{t+1})$
  \item $\beta_t(i)$: $\beta_t(i) = Pr(o_{t+1:T} |Q_t = S_i, \lambda)$ with calculation: $\beta_t(i) = \sum_{j=1}^N A_{ij} b_j (o_{t+1})\beta_{t+1}(j)$
  \item $\gamma_t(i)$: $\gamma_{t}(i)=Pr(Q_t = S_i|o_{1:T} , \lambda) $
  \item $\xi_t(i, j) = Pr(Q_t = S_i, Q_{t+1} = S_j |o_{1:T} , \lambda). $
\end{itemize}

The $\xi$ variable involves three other values: t (the time) and (i, j) which are
state indexes. Comparing the definition of $\gamma$ and $\xi$, we immediately get (by the law
of total probability):

\begin{equation}
\gamma_{t}(i)= \sum_{j=1}^{N}\xi_t(i, j) 
 \end{equation}

The parameters $\lambda$ = ($\pi , A , B$ ) can be updated using $\gamma$ and $\xi$.
Using the definition of conditional probabilities, we have
\begin{equation}
\xi_t(i, j) Pr(o_{1:T} |\lambda) = Pr(Q_t = S_i, Q_{t+1} = S_j , o_{1:T} |\lambda).
\end{equation}

we can find the probability \\$Pr(Q_t = S_i, Q_{t+1} = S_j , o_{1:T} |\lambda)$ and use
it to compute $\xi_t(i, j)$. This probability can be factored into the product of
four probabilities: $\alpha_t(i), A_{ij}$ , $b_j (o_{t+1}) $ and $ \beta_{t+1}(j)$

\begin{equation}
\xi_t(i, j)=\frac{\alpha_t(i) A_{ij} b_j (o_{t+1}) \beta_{t+1}(j)}{Pr(o_{1:T}|\lambda)}
\end{equation}

The entire training algorithm are shown in Algorithm \ref{code1}.

\begin{algorithm}[h]  
\caption{Training Algorithm}\label{code1}  
\begin{algorithmic}[1]
\STATE Initialize the parameters $\lambda^{(1)}$ (e.g., randomly)
\STATE $\tau \gets 1$   
\WHILE {the likelihood has not converged}  
\STATE Use the forward procedure to compute $\alpha_t$(i) for all t (1 $\leq$ t $\leq$ T) and all
i (1 $\leq$ i $\leq$ N) based on $\lambda^{( \tau)}$
\STATE Use the backward procedure to compute $\beta_t$(i) for all t (1 $\leq$ t $\leq$ T) and
all i (1 $\leq$ i $\leq$ N) based on $\lambda^{(\tau)}$
\STATE Compute $\gamma_t$(i) for all t (1 $\leq$ t $\leq$ T) and all i (1 $\leq$ i $\leq$ N) according to
the equation in Table 1
\STATE Compute $\xi_t$(i, j) for all t (1 $\leq$ t $\leq$ T − 1) and all i, j (1 $\leq$ i, j $\leq$ N)
according to the equation in Table 1
\STATE Update the parameters to $\lambda^{(r+1)}$ 
\begin{equation}
\pi_i^{(\tau+1)}=\gamma_1(i)
\end{equation}  
\begin{equation}
A_{ij}^{(\tau+1)}=\frac{\sum_{t=1}^{T-1} \xi_t(i, j)}{\sum_{t=1}^{T-1} \gamma_t (i)}
\end{equation}
\begin{equation}
b_{j}^{(\tau+1)}(k)=\frac{\sum_{t=1}^{T}\Vert o_t=k \Vert \gamma_t(j)}{\sum_{t=1}^{T} \gamma_t (j)}
\end{equation}
\STATE $\tau \gets \tau +1 $  
\ENDWHILE   
\end{algorithmic}  
\end{algorithm}

\subsection{Real time decoding}
The aim is to find the most probable hidden state sequence that could generate the observations sequence
 produced by hearing the soloist’s playing. In the score followers developed during this project, a revised Viterbi algorithm is used to find out which state the soloist is most likely to be in (given the sequence of observations of what notes the soloist has most recently played).

Implemented in the traditional fashion\cite{rabiner1989tutorial}, this algorithm finds the globally optimum path through the Hidden Markov Model states to the most probable current state, using
the history of observations seen. But this causes huge computational complexity and the system cannot be used in practice
\footnote{Most classical musical pieces have $O(100 - 10000)$ chords. For example, the solo piano part of Rachmaninoff’s piano concerto No. 3 d-moll has N $\simeq$ 5000 chords only in the first movement.}.  

Although one might consider some pruning techniques to reduce computational complexity,
pruning is not valid within the context of handling arbitrary skips since skips rarely
occur compared to other state transitions. Therefore, it seems necessary to introduce some
constraints to the performance HMM.

The problem with large computational complexity arises from the non-zero values of the transition probability $a_{ij}$ for large $|i-j|$. Here, we assume the transition probability can be summarised as

\begin{equation}
a_{ij} = \tilde{A_{ij}}  + \mu.
\end{equation}

where $\tilde{A_{ij}}$ is a band matrix satisfying $\tilde{A_{ij}}$= $A_{ij}$ when $i - W_1 \leq j \leq i + W_2$, otherwise $\tilde{A_{ij}}$ = 0, which describes transitions within neighbouring states.
$\mu$ is a prior distribution got from training part \cite{couasnon1994using}  depicting arbitrary repeats/skips\footnote{performers are likely to resume their performance from the beginning of a sentence/phrase when they make mistakes \cite{nakamura2016real}}.

\begin{equation}
a_{ij}=\mu, \qquad for \quad j<i-W_1 \quad or \quad j > i+W_2
\end{equation}
where $W_1$ and $W_2$ are small positive integers which define a neighbourhood of states.

%%%%%%%%%%%%%
Given an observation sequence $o_{1:T}$, we use a new variable $\delta$, defined by Equation (\ref{eq1}), to find the best path. $W$ is the sliding window width and $W=W_1+W_2+1$. $\delta$ has recursive relationship, as shown in equation (\ref{eq2})
\begin{equation}
\delta_t(i) = \max_{Q_{1:t-1}} Pr(Q_{1:t-1}, o_{1:t}, Q_t = S_i|\lambda)
\label{eq1}
\end{equation}

\begin{equation}
\delta_{t+1}(i) = \max_{1\leq j\leq N} (\delta_t(j) A_{ji} b_i(o_{t+1}))
\label{eq2}
\end{equation}

%repeat and error

\begin{algorithm}[h]  
\caption{Decoding Algorithm}  
\begin{algorithmic}[1]
\STATE \textbf{Initialization}: $\delta_1{(i)}=\pi_ib_i(o_1)$,$\psi_1(i)=0$ for all 1 $\leq$ i $\leq$ N 
\STATE \textbf{Slide algorithm window}: 
\STATE \textbf{Forward recursion}: For t = 1,2, $\cdots$ , $T-2$, $T-1$ and all 1 $\leq$ i $\leq$ N
\begin{equation}
\delta_{t+1}{(i)}=max (\delta_t{(j)} a_{ji} b_i(o_{t+1})) \label{decodingeq}
\end{equation}  
\begin{equation}
\psi_{t+1}{(i)}= arg max (\delta_t{(j)} a_{ji} )
\end{equation}
\STATE \textbf{Output}:  The optimal state $q_T$ is determined by
\begin{equation}
q_T=arg \max_{1\leq i \leq N} \delta_{t+1}{(i)}
\end{equation}    
\label{code2}  
\end{algorithmic}  
\end{algorithm}  

\subsection*{Theoretical evaluation}
We can rewrite equation (\ref{decodingeq}) as below:
\begin{equation}
\delta_{t+1}{(i)} = b_i(o_{t+1}) max \lbrace \max_{j \in nbh(i)}[ \delta_t{(j)} A_{ji}], \max_{j}[ \delta_t{(j)} \mu] \rbrace \label{decodingeq2}
\end{equation}
\vspace{0.03in}

where nbh(i) = $\lbrace j|j - W_1 \leq i \leq j + W_2 \rbrace$ denotes the set of neighbouring states of
i. Since the factor $ \max_{j}[ \delta_t{(j)} \mu]$ in the last equation is independent of i and can be calculated with $O(N)$ complexity, the decoding algorithm expression has $O(WN)$ computation complexity compared with previous $O(N^2)$ complexity. Therefore, a fast Viterbi algorithm can
be used efficiently for the HMM if W $\ll$ N.

\subsection{Two hands parallel HMM}
We construct a two hands parallel HMM, with each hand as part HMMs corresponding to the HMM described above. The two then merged their outputs, assuming there is no hand crossing in performance.

The two part HMMs transits and outputs an observed symbol at each time. The
state space of the parallel HMM is given as a triplet
$k = (\eta, f_L, f_R)$ of the hand information, where $\eta$  indicate which of the HMMs works, and $f_L$ and $f_R$ indicate the current states of the
part HMMs. \cite{nakamura2014merged}

\section{Accompanist}

A soloist will naturally incorporate expressive features
in the playing, involving shaping of the tempo and
intensity of the playing in ways not explicitly represented
in the score. 

A human accompanist would not wait for every note to
be played by the soloist before playing accompaniment.
Instead they anticipate that the soloist will move onto the
next note in the score and play the appropriate
accompaniment, then use the incoming information
from the soloist to update their belief of where the
soloist is in the score and adjust their accompaniment if
necessary.

In a similar fashion, this system uses the Hidden
Markov Model representation to work out what the
next sequential state is, playing the accompaniment for
that state at the time it expects the next state to occur.
As it receives and processes the soloist’s actual input
and locates the HMM state that the soloist has
actually reached, it adjusts the accompaniment if
necessary.

\subsection{Beat tracking}
We implement beat tracking to monitor the performance's tempo and provide a reference to the accompaniment speed.

The system incorporates a simple version of beat
tracking. This allows small tempo fluctuations to be
tracked, and the soloist’s output to be anticipated in a
timely fashion. Modelling the score by temporal units
assisted us greatly with including beat tracking in the
accompanist. Our implementation was simpler than \cite{dixon2001automatic} but
was effective.

The accompanist used an internal tempo measure that
was continually adjusted to match the soloist’s estimated
current tempo, using a local window of notes recently
played by the soloist and measuring the time in between
those notes (relative to the notes’ expected durations). If
the soloist is currently judged to be in a ghost state (i.e.
they have deviated from the score), then the last input is
not considered as valid for use in updating the tempo. If,
though, the soloist is currently judged to be in a normal
state (i.e. they can be found on the score), then the score
follower works out how long the previous note should
have been and compares this with the actual length of the
last note. The current tempo is based on an average of
the recent (valid) tempo observations. The largest and
smallest tempo observations are ignored and a mean is
taken of the remaining tempo observations, to generate
an estimate of the current tempo.

\subsection{Controlling dynamics of the performance}
The system can track the volume of the soloist’s playing
using MIDI information and replicate that volume in the
dynamic level of the accompaniment output, playing
the accompaniment at a very slightly lower volume than
the soloist. In this way the system allows the soloist line
to be prominent but also matches the dynamic markings
of their playing. We felt it was more important to be
responsive to the soloist’s dynamic interpretations than
to allow the accompanist to play at a dynamic marking
independent of the soloist’s dynamics.

\subsection{Rule-based Reactive accompanist}
If a human accompanist hears their soloist deviate slightly from the score, it takes
time for the accompanist to relocate the soloist and adjust their playing from the expected
accompaniment to the accompaniment matching the soloist.

It would be reasonable to have the computer accompanist only respond to a deviation
on the next state after a deviation from the score was identified: replicating the
slight delay that a human accompanist would also have. This is on the assumption that
the states are modelled such that they are close enough together in timing for the delay
not to be too noticeable.

We use a musical accompanying rule to generate our accompaniment\cite{friberg1991generative}, i.e. a chord match with some certain chords.

\begin{acknowledgments}
Thanks to the support of the University of Hong Kong Computer Science Department summer research internship programme.
\end{acknowledgments} 

%%%%%%%%%%%%%%%%%%%%%%%%%%%%%%%%%%%%%%%%%%%%%%%%%%%%%%%%%%%%%%%%%%%%%%%%%%%%%
%bibliography here

\bibliography{smc2017template}

\end{document}